\documentstyle{article}

\begin{document}

\title{Piecing Together the Biggest Puzzle of All}

\author{MARTIN J. REES\\\\
Institute of Astronomy, University of Cambridge\\
Madingley Road,\\ Cambridge CB3 0HA\\
 United Kingdom\\
Email: mjr@ast.cam.ac.uk}
\date{\phantom{-}}


\maketitle

Throughout human history, our existence and our place in nature have
been enduring mysteries. Only during the 20th century have astronomers
and cosmologists fully realized the scale of our cosmos and understood
the physical laws that govern it. This story sets our Earth in an
evolutionary context stretching back before the birth of our solar
system -- right back, indeed, to the primordial event that set our entire
cosmos expanding about 12 billion years ago. Quasars, black holes,
neutron stars, and the big bang have entered the general vocabulary,
if not the common understanding. Fundamental questions about our
universe that were formerly in the realm of speculation are now within
the framework of empirical science.

\section {   Gravity, General Relativity, Neutron Stars, and Black Holes} 

Central to many of these questions is gravity. It's the governing
force in the cosmos. It holds planets in their orbits, binds stars and
galaxies, and determines the fate of our universe. Isaac Newton's 17th
century theoretical description of gravity remains accurate enough to
program the trajectories of spacecraft on their journeys to Mars,
Jupiter, and beyond. But ever since 1905, when Albert Einstein's
special theory of relativity showed that instantaneous transfer of
information was precluded, physicists accepted that Newton's laws
would be inadequate when the motions induced by gravity approached the
speed of light. Einstein's general relativity (published in 1916),
however, copes quite consistently even with situations when gravity is
overwhelmingly strong.  

General relativity ranks as one of the two
pillars of 20th century physics; the other is quantum theory, a
conceptual revolution that presaged our modern understanding of atoms
and their nuclei. Einstein's intellectual feat was especially
astonishing because, unlike the pioneers of quantum theory, he wasn't
motivated by any experimental enigma.

It took 50 years before astronomers discovered objects whose gravity
was strong enough to manifest the most distinctive and dramatic
features of Einstein's theory. In the early 1960s, ultraluminous
objects (quasars) were detected. They seemed to require an even more
efficient power supply than nuclear fusion, the process by which stars
shine; gravitational collapse seemed the most attractive
explanation. The American theorist Thomas Gold expressed the
exhilaration of theorists at that time. In an after-dinner speech at
the first big conference on the new subject of relativistic
astrophysics, held in Dallas in 1963, he said, ``The relativists with
their sophisticated work [are] not only magnificent cultural ornaments
but might actually be useful to science! Everyone is pleased: the
relativists, who feel they are being appreciated, who are suddenly
experts in a field they hardly knew existed; the astrophysicists for
having enlarged their domain. ... It is all very pleasing, so let us
hope it is right."

Observation-using the novel techniques of radio and X-ray
astronomy-bore out Gold's optimism. In the 1950s, the world's best
optical telescopes were concentrated in the United States,
particularly in California. This shift from Europe had come about for
climatic as well as financial reasons. However, radio waves from space
can pass through clouds, so Europe (and Australia) were able to
develop the new science of radio astronomy without any climatic
handicap.

Some of the strongest sources of cosmic radio noise could be readily
identified. One was the Crab Nebula, the expanding debris of a
supernova explosion witnessed by oriental astronomers in
A.D. 1054. Other sources were remote extragalactic objects, involving,
we now realize, energy generation around gigantic black holes. These
detections were unexpected. The physical processes that emit the radio
waves, although now well understood, were not predicted.  

The most
remarkable serendipitous achievement of radio astronomy was the
discovery of neutron stars in 1967 by Anthony Hewish and Jocelyn
Bell. These stars are the dense remnants left behind at the core of
some supernova explosions. They were detected as pulsars: They spin
(sometimes many times per second) and emit an intense beam of radio
waves that sweeps across our line of sight once per revolution. The
importance of neutron stars lies in their extreme physics-colossal
densities, strong magnetic fields, and intense gravity.  

In 1969, a
very fast (30-Hz) pulsar was found at the center of the Crab
Nebula. Careful observations showed that the pulse rate was gradually
slowing. This was natural if energy stored in the star's spin was
being gradually converted into a wind of particles, which keeps the
nebula shining in blue light. Interestingly, this pulsar's repetition
rate, 30 per second, is so high that the eye sees it as a steady
source. Had it been equally bright but spinning more slowly-say, 10
times a second-the remarkable properties of this little star could
have been discovered 70 years ago. How would the course of 20th
century physics have been changed if superdense matter had been
detected in the 1920s, before neutrons were discovered on Earth? One
cannot guess, except that astronomy's importance for fundamental
physics would surely have been recognized far sooner.

Neutron stars were found by accident. No one expected them to emit
strong and distinctive radio pulses. If theorists in the early 1960s
had been asked the best way to detect a neutron star, most would have
suggested a search for x-ray emission. After all, if neutron stars
radiate as much energy as ordinary stars, but from a much smaller
surface, they must be hot enough that the radiation from them is in
X-rays. So it was x-ray astronomers who appeared best placed to
discover neutron stars.

X-rays from cosmic objects get absorbed by Earth's atmosphere,
however, and so can only be observed from space.  X-ray astronomy,
like radio astronomy, received its impetus from wartime technologies
and expertise. In this case it was scientists in the United States who
took the lead, especially the late Herbert Friedman and his colleagues
at the U.S. Naval Research Laboratory. Their first x-ray detectors,
mounted on rockets, each yielded only a few minutes of useful data
before they crashed back to the ground. X-ray astronomy spurted
forward in 1970 when NASA launched the first x-ray satellite, which
could gather data for years at a time. Through this project and its
many successors, x-ray astronomy has proved itself to be a crucial new
window on the universe.  

X-rays are emitted by unusually hot gas and
by especially intense sources. X-ray maps of the sky consequently
highlight the hottest and most energetic objects in the
cosmos. Neutron stars, which pack at least as much mass as the sun in
a volume little more than 10 kilometers in diameter, are among
these. Their gravity is so strong that relativistic corrections amount
to 30\%.  

Some stellar remnants, we now suspect, collapse beyond
neutron-star densities to form black holes, which distort space and
time even more than a neutron star does. An astronaut who ventured
within the horizon around a black hole could not transmit any light
signals to the external world-it is as though space itself is being
sucked inward faster than light moves out through it. An external
observer would never witness the astronaut's final fate: Any clock
would appear to run slower and slower as it fell inward, so the
astronaut would appear impaled at the horizon, frozen in time.

The Russian theorists Yakov Zeldovich and Igor Novikov, who studied
how time was distorted near collapsed objects, coined the term ``frozen
stars" in the early 1960s. The term ``black hole" was coined in 1968,
when John Wheeler described how ``light and particles incident from
outside ... go down the black hole only to add to its mass and
increase its gravitational attraction."  

The black holes that
represent the final evolutionary state of stars have radii of 10 to 50
kilometers. But there is now compelling evidence that holes weighing
as much as millions, or even billions, of suns exist at the centers of
most galaxies. Some of them manifest themselves as
quasars-concentrations of energy that outshine all the stars in their
host galaxy-or as intense sources of cosmic radio emission. Others,
including one at our own galactic center, are quiescent, but affect
the orbits of stars passing close to them.  

Viewed from outside, black
holes are standardized objects: No traces persist to distinguish how a
particular hole formed, nor what objects it swallowed. In 1963 the New
Zealander Roy Kerr discovered a solution of Einstein's equations that
represented a collapsed rotating object. The ``Kerr solution" acquired
paramount importance when theorists realized that it describes
space-time around any black hole. A collapsing object quickly settles
down to a standardized stationary state characterized by just two
numbers: those that measure its mass and its spin. Roger Penrose, the
mathematical physicist who perhaps did the most to stimulate the
renaissance in relativity theory in the 1960s, has remarked, ``It is
ironic that the astrophysical object which is strangest and least
familiar, the black hole, should be the one for which our theoretical
picture is most complete."

The discovery of black holes opened the way to testing the most
remarkable consequences of Einstein's theory. The radiation from such
objects comes primarily from hot gas swirling downward into the
``gravitational pit." It displays huge Doppler effects, as well as
having an extra redshift because of the strong gravity. Spectroscopic
study of this radiation, especially the x-rays, can probe the flow
very close to the hole and diagnose whether the shape of space near it
agrees with what theory predicts.

\section{The Expanding Cosmos} 

Our own Milky Way contains about 100 billion
stars, mainly in a disc orbiting a central hub. Nothing was known
about any more remote parts of the universe until the 1920s, but it's
now recognized that our galaxy is one among billions.  

Galaxies occur
mostly in groups or clusters, held together by gravity. Our own Local
Group, a few million light-years across, contains the Milky Way and
Andromeda, together with 34 smaller galaxies. This group is near the
edge of the Virgo Cluster, an archipelago of several hundred galaxies,
whose core lies about 50 million light-years away. Clusters and groups
are themselves organized in still larger aggregates. The so-called
Great Wall, a sheetlike array of galaxies about 200 million
light-years away, is the nearest and most prominent of these giant
features.

Perhaps the most important single broad-brush fact about our universe
is that all galaxies (except for a few nearby galaxies in our own
cluster) are receding from us. Moreover, the redshift-a measure of the
recession speed-is larger for the fainter and more distant
galaxies. We seem to be in an expanding universe where clusters of
galaxies get more widely separated-more thinly separated through
space-as time goes on.

The simple relation between redshift and distance is named after Edwin
Hubble, who first claimed such a law in 1929. Hubble could only study
relatively nearby galaxies, whose recession speeds were less than 1\%
of the speed of light. Thanks to technical advances and larger
telescopes, the data now extend to galaxies whose apparent recession
amounts to a good fraction of the speed of light. It is conceptually
preferable to attribute the redshift to ``stretching" of space while
the light travels through it. The amount of redshifting-in other
words, the amount the wavelengths are stretched-tells us how much the
universe has expanded while the light has been traveling toward us.

Models for an expanding, homogeneous universe, some based on
Einstein's general relativity, had been devised in the 1920s and
1930s. But there was then little quantitative evidence on the extent
to which our universe was actually homogeneous. Still less was it
possible to discriminate between alternative models.

Astronomers are now mapping many more clusters like Virgo and more
features like the Great Wall. But deeper surveys don't seem to reveal
anything even larger. A box 200 million light-years on a side (a
distance still small compared to the horizon of our observations,
which is about 10 billion light-years away) can accommodate the
largest aggregates. Such a box, wherever it was placed in the
universe, would contain roughly the same number of galaxies, grouped
in a statistically similar way, into clusters, filamentary structures,
and so on.  

Even the very biggest conspicuous cosmic structures are
small compared with the largest distances our telescopes can
probe. This makes the science of cosmology possible, by allowing us to
define the average properties of our universe and to use simple,
homogeneous models as a valid approximation.

In the 1950s, Allan Sandage was a lone pioneer in advocating how the
200-inch (5-m) Mount Palomar telescope could probe deep enough into
space and, therefore, far enough back in the past, to test
cosmological models. To detect changes in the expansion rate, or
evolution in the galaxy population, it is necessary to look at objects
so far away that their light set out billions of years ago.  

In the
last 40 years, the development of ever more capable and revealing
telescopes and observational techniques has made this possible. Over a
dozen telescopes with 4-meter mirrors were built during the 1970s and
1980s. Replacement of photographic plates, with 1\% quantum efficiency,
by solid-state detectors with efficiencies up to 80\% hugely enhanced
detectability of faint and distant objects. A new generation of still
larger telescopes (of which the two Keck Telescopes in Hawaii are the
first) is now coming into service. Perhaps most impressive of all is
the unimaginatively named Very Large Telescope, a cluster of four
telescopes, each with an 8.2-meter mirror, constructed in the Chilean
Andes by a consortium of European nations. This instrument not only
collects more light than any previous telescope but is also intended
to yield sharper images, by compensating for the fluctuations in the
atmosphere and linking the telescopes together so they can function as
an interferometer.

There also have been dramatic advances in space-based
observations. Although initially dogged by delays, flaws, and cost
overruns, the Hubble Space Telescope has been fulfilling the hopes
astronomers had for it. The Hubble Deep Field images-obtained by
pointing steadily for several days at a small patch of sky-reveal
literally hundreds of faint smudges, even within a field of view so
small that it would cover less than 1/100 the area of the full
moon. Each smudge is an entire galaxy, thousands of light-years in
size, which appears so small and faint because of its huge
distance. We are viewing these remote galaxies at a very primitive
evolutionary stage. They have as yet no complex chemistry: There would
have been very little oxygen, carbon, and other elements to make
planets, and so scant chance of life.  

We now have snapshots that take
us billions of years back in time, to the era when the first galaxies
were forming. The first stars may actually have formed even earlier,
in aggregates smaller than present-day galaxies, which are too faint
for even the largest existing telescopes to reveal.

\section {   ``Fossils" of the Hot Beginning}

 What about still more remote epochs,
before even the first stars had formed? In the later 1920s, the
MIT-trained Belgian priest Georges Lemaître, along with Aleksandr
Friedmann in Russia, was a pioneer of the idea that everything began
in a dense state and that its structure unfolded as it
expanded. Lemaître wrote: ``The evolution of the universe can be
likened to a display of fireworks that has just ended: some few wisps,
ashes and smoke. Standing on a well-chilled cinder we see the fading
of the suns, and try to recall the vanished brilliance of the origin
of the worlds."  

This ``vanished brilliance" was revealed in 1965. Arno
Penzias and Robert Wilson, two scientists at the Bell Telephone
Laboratories striving to reduce the noise in an antenna in Holmdel,
New Jersey, serendipitously discovered that all space is slightly
warmed by microwaves with no apparent source. In 1990, John Mather and
his colleagues, using NASA's Cosmic Background Explorer (COBE)
satellite, showed that the spectrum obeys a ``blackbody" or thermal
law, with a precision of one part in 10,000-just what would be
expected if it were indeed a relic of a ``fireball" phase when
everything in our universe was squeezed intensely hot, dense, and
opaque. The cosmic expansion would have cooled and diluted the
original radiation and stretched its wavelength, but it would still be
around, pervading all of space. 

 The present background temperature is
only 2.728 degrees above absolute zero, but this represents a
surprising amount of heat: 412 million quanta of radiation (photons)
in each cubic meter of the present-day universe. In contrast, all the
observed stars and gas in the universe, if spread uniformly through
space, would amount to only about 0.2 atoms per cubic meter-more than
a billion times less than the photon density.

According to the big bang theory, everything would once have been
compressed hotter than the center of a star-certainly hot enough for
nuclear reactions. The most important of these reactions happen at a
temperature of roughly a billion degrees. However, the universe cooled
below this temperature within 3 minutes, and (fortunately for us) this
didn't allow time to process primordial material into iron, as in the
hottest stars-nor even into carbon, oxygen, etc.

This was contrary to George Gamow's conjecture that the entire
Periodic Table was "cooked" in the early universe. In the 1950s, Fred
Hoyle, William Fowler, and Geoffrey and Margaret Burbidge-and, in
parallel work, Alistair Cameron-developed an alternative scheme, which
quantitatively explained almost the entire Periodic Table as the
outcome of nuclear fusion in stars and supernovae. After later
refinements, the calculated mix of atoms is gratifyingly close to the
proportions now observed.  

The oldest stars, which would have formed
from gas early in galactic history, when it was less ``polluted," are
indeed deficient in heavy elements, just as stellar nucleosynthesis
theory would lead one to expect. However, even the oldest objects
turned out to be 23\% to 24\% helium: No star, galaxy, or nebula has
been found where helium is less abundant than this. It seems as though
the galaxy started not as pure hydrogen but was already a mix of
hydrogen and helium.

The ``hot big bang" theory neatly solves this mystery. Reactions in the
hot early phases would turn about 23\% of the hydrogen into helium, but
the universe cooled down so fast that there wasn't time to synthesize
the elements higher up the Periodic Table (apart from a trace of
lithium). Attributing most cosmic helium to the big bang thus solved a
long-standing problem-why there is so much of it, and why it is so
uniform in its abundance-and emboldened cosmologists to take the first
few seconds of cosmic history seriously.  

Another product of the big
bang is deuterium (heavy hydrogen). Deuterium's abundance relative to
hydrogen was until recently uncertain. But the ratio, measured in
Jupiter, in interstellar gas, and in remote intergalactic clouds, is
now pinned down to be about 1/50,000. The origin of even this trace
poses a problem, because deuterium is destroyed rather than created in
stars. As a nuclear fuel, it is easier to ignite than ordinary
hydrogen, so newly formed stars destroy their deuterium during their
initial contraction, before settling down in their long,
hydrogen-burning phases.

If we assume a present universe-wide average density of 0.2 atoms per
cubic meter and compute what mixture of atoms would emerge from the
cooling fireball, we find that the proportions of hydrogen, deuterium,
and helium (and, as a bonus, lithium as well) agree with
observations. This is gratifying, because the observed abundances
could have been entirely out of line with the predictions of any big
bang; or they might have been consistent, but only for a density that
was far below, or else far above, the range allowed by observation.

\section {Should We Believe the Big Bang Scenario?}

The extrapolation by astrophysicists and cosmologists back to a stage
when the universe had been expanding for a few seconds deserves to be
taken as seriously as, for instance, what geologists or
paleontologists tell us about the early history of our Earth: Their
inferences are just as indirect and generally less quantitative.

Moreover, there are several discoveries that might have been made over
the last 30 years which would have invalidated the big bang hypothesis
and which have not been made -- the big bang theory has lived dangerously
for decades and survived. Here are some of those absent observations:

\begin{itemize}
\item
Astronomers might have found an object whose helium abundance was
far below the amount predicted from the big bang -- 23\%. This would have
been fatal, because extra helium made in stars can readily boost
helium above its pregalactic abundance, but there seems no way of
converting all the helium back to hydrogen.

\item The background radiation measured so accurately by the Cosmic
Background Explorer satellite might have turned out to have a spectrum
that differed from the expected ``blackbody" or thermal form. What's
more, the radiation temperature could have been so smooth over the
whole sky that it was incompatible with the fluctuations needed to
give rise to present-day structures like clusters of galaxies.

\item A stable neutrino might have been discovered to have a mass in
the range of 100 to 106 electron volts. This would have been fatal,
because the hot early universe would have contained almost as many
neutrinos as photons. If each neutrino weighed even a millionth as
much as an atom, they would, in toto, contribute too much mass to the
present universe-more, even, than could be hidden in dark
matter. Experimental physicists have been trying hard to measure
neutrino masses, but they are seemingly too small to be important
contributors to dark matter.  

\item The deuterium abundance could have been so high that it was
inconsistent with big bang nucleosynthesis (or implied an unacceptably
low baryon density).
\end{itemize}

The big bang theory's survival gives us confidence in extrapolating
right back to the first few seconds of cosmic history and assuming
that the laws of microphysics were the same then as now.

\section {   The Emergence of Structure}

 If our universe started off as a hot,
amorphous fireball, how did it differentiate into the observed pattern
of stars, galaxies, and clusters? This is actually a natural outcome
of the workings of gravity, which leverages, over time, even very
slight initial irregularities into conspicuous density contrasts.

Theorists can now follow virtual universes in a computer. Slight
fluctuations are fed in at the start of the simulation. As the
universe expands, incipient galaxies and larger structures emerge and
evolve. The purely gravitational aspects of this process can be
modeled fairly well. However, when gas contracts under gravity into a
protogalaxy, its density has to rise by many powers of 10 before stars
form, and complicated dynamics and radiative transfer determine what
their masses are. Moreover, the energy output from the first stars
exerts uncertain feedback on what happens later. This is too
complicated to be computed and has to be approximated by plausible
``recipes," chosen in the light of local observations.

Despite these limitations, simulations of structure formation have
achieved remarkable success in accounting for present-day morphology
of galaxies and clusters. Moreover, these models can be checked by
seeing how well they account also for the new high-redshift data,
which tell us what the universe was like at earlier times.

There is another consistency check on computed scenarios. In these
simulations, the predicted sizes and clustering of galaxies depend on
the form and amplitude of the initial fluctuations. The microwave
background should bear the imprint of these fluctuations and thus
offers an independent line of evidence on their amplitude. This
radiation in effect comes from a surface so far away that it is being
observed at an era when the fluctuations were still of small
amplitude. Radiation from an incipient cluster on that surface would
appear slightly cooler, because it loses extra energy climbing out of
the gravitational pull of an overdense region; conversely, radiation
from the direction of an incipient void would be slightly hotter. The
predicted fractional differences in the temperature over the sky due
to this effect are about one part in $10^5$. On some scales, a somewhat
larger Doppler fluctuation is predicted, due to the associated
motions. 

 Measuring one part in 100,000 of background radiation, which
is itself 100 times cooler than the atmosphere, is a daunting
technical challenge. But it has now been achieved. Fluctuations were
first measured by George Smoot and his colleagues, using 4 years of
data gathered by the COBE satellite. These measurements were
restricted to angular scales exceeding 7 degrees, however. They since
have been supplemented and extended by ground-based and balloon
experiments. The amplitudes are indeed consistent with what is
required for galaxy formation. Within a few years, two new
spacecraft -- NASA's Microwave Anisotropy Probe and the European Space
Agency's Planck satellite -- will yield data precise enough to settle
many key questions about cosmology, the early universe, and how
galaxies emerged.

\section {   Dark Matter, Omega, and the First Microsecond}

 In about 5 billion
years, the sun will die and Earth with it. At about the same time
(give or take a billion years) the Andromeda galaxy, already falling
toward us, may crash into our own Milky Way. But will the universe go
on expanding forever? Or will the entire firmament eventually
recollapse to a big crunch, where everything will suffer the same fate
as an unwary astronaut who falls into a black hole?  

The answer
depends on how much the cosmic expansion is being decelerated by the
gravitational pull that everything exerts on everything else. It is
straightforward to calculate that the expansion can eventually be
reversed if there is, on average, more than about five baryons in each
cubic meter -- the so-called critical density. (A baryon is the
collective term for protons and neutrons, the heavy ingredients of all
atoms.) That doesn't sound like much. But if all the galaxies were
dismantled, and their constituent stars and gas spread uniformly
through space, they'd make an even emptier vacuum -- one baryon in every
10 cubic meters. Add to that what seems to be a similar amount of
material in diffuse intergalactic gas, and the resulting density
amounts to 0.2 baryons per cubic meter.  

That's 25 times less than the
critical density, which at first sight implies perpetual
expansion. But the actual situation is less
straightforward. Astronomers have discovered that galaxies, and even
entire galaxy clusters, would fly apart unless held together by the
gravitational pull of five to 10 times more material than we actually
see. This is the famous ``dark matter" mystery.  

There are many
candidates for dark matter. Earlier ideas included very faint stars
(known as ``brown dwarfs"), or the remnants of massive stars. However,
most cosmologists suspect that dark matter is mainly exotic particles
left over from the big bang and is not made of baryons at all.

There are two main reasons for this belief. First, the proportions of
helium and deuterium that are calculated to emerge from the big bang
are sensitive to the baryon density and would be discrepant with
observations if the average baryonic density were, say, between 1 and
2, rather than 0.2 per cubic meter. Extra dark matter in exotic
particles that do not participate in nuclear reactions, however, would
not scupper the concordance with a baryon density of only 0.2 per
cubic meter.

Second, the formation of galaxies would be hard to understand if the
bulk of their mass were baryonic. Nonbaryonic matter can cluster more
efficiently in the early universe because it does not feel the
opposing influence of radiation pressure. If the universe were purely
baryonic, it would be hard to reconcile its present highly structured
state with the small amplitude of the primordial fluctuations implied
by the microwave background anisotropies.

The dark matter mystery may yield to a three-pronged attack: 

\begin{enumerate}
\item{\it Direct
detection}.  Using sensitive detectors in underground laboratories,
searches are now under way for dark matter candidates, including heavy
neutral particles and axions.  

\item
 {\it Progress in particle physics}. If we
knew more about the types of particles that could exist in the
ultraearly universe, then we could confidently calculate how many
should survive from the first microseconds of the big bang and how
much dark mass they would contribute. 

\item {\it Simulations of galaxy
formation and large-scale structure}. When and how galaxies form and
cluster depends on what their gravitationally dominant constituent
is. It is possible to simulate the formation of galaxies on a
computer, making alternative assumptions about the dark matter. If one
assumption yielded an outcome that matched real galaxies especially
closely, this would at least be corroborative evidence for one
candidate rather than another.
\end{enumerate}

Cosmologists denote the ratio of the actual density to the critical
density by $\Omega$. There is certainly enough dark matter to make $\Omega =
0.2$. Until recently, we couldn't rule out several times this
amount -- comprising the full critical density, $\Omega  = 1$ -- in the space
between clusters of galaxies. But it now seems that, in toto, atoms
and dark matter don't contribute more than about 30\% of critical
density.  

We can never be sure of the long-term future: New physics
may intervene, and domains beyond our observational range may be
different from the part of the universe we can see. But with those
provisos, the odds favor perpetual expansion. The galaxies will
disperse, and they will fade as their stars all die and their material
gets locked up in old white dwarfs, neutron stars, and black holes.

There is, moreover, tantalizing evidence for an extra repulsion force
that overwhelms gravity on the cosmic scale. Studies of the redshift
versus the apparent brightness of distant supernovae suggest that
galaxies may disperse at an accelerating rate. Science rated
this
-- perhaps prematurely -- as the most important discovery of 1998, in
any field. If this work is corroborated, the forecast is an even
emptier universe. All galaxies beyond our local group will accelerate
away, disappearing completely from view as their redshift rises
exponentially toward infinity.

The idea of a cosmological repulsion goes back to 1917, when Einstein
introduced an extra term into his equations, which he called the
cosmological constant, or $\lambda$. His motive was to allow a static
universe, in which the repulsive force counteracted gravity. He later
abandoned the idea, calling it his ``biggest blunder," when Hubble
discovered that the universe was actually expanding. However, from our
modern perspective, $\lambda$  can be envisaged as dark energy latent in empty
space. It leads to a repulsion because, according to Einstein's
equation, gravity depends on pressure as well as density, and if the
pressure is sufficiently negative (as it has to be for vacuum energy),
the net effect is repulsive.*

The cosmological constant corresponds to a vacuum energy that is
unchanging as the universe expands. Cosmologists have recently
suggested variants -- forms of dark energy, dubbed quintessence, with
negative pressure that could gradually decay.  

There is another line
of evidence for some form of dark energy, quite independent of the
indications from supernovae for an accelerating universe. Theory tells
us that fluctuations or ripples in the microwave background should be
biggest on a particular length scale that is related to the maximum
distance a sound wave can travel. The angular scale corresponding to
this length depends, however, on the geometry of the universe.

Measurements have now pinned down the angular scale of this Doppler
peak with better than 10\% precision. The results are consistent with a
``flat" universe. In contrast, if there were no mass-energy apart from
enough baryons and dark matter to yield 0.2 to 0.3 for $\Omega$, we would be
in an open universe, where this angle would be smaller by a factor of
2 -- definitely in conflict with observations. Reconciliation with the
microwave background measurements would be achieved if dark energy
made up the balance, so that the universe was, after all, flat; the
dominant dark energy would then drive the accelerating expansion.

Within just the last 2 years, a remarkable concordance among several
seemingly independent observations has emerged, leading to a preferred
choice for the key parameters describing our universe. It seems that
the universe is flat, with baryons providing 4\% of the mass-energy,
dark matter 20\% to 30\%, and dark energy the rest, i.e., 66\% to
76\%. The flatness vindicates a natural prediction of the ``inflationary
theory" (discussed below). The expansion accelerates because dark
energy (with negative pressure) is the dominant constituent. But there
seems nothing natural about the actual split between the three
different contributions.  

Is there any explanation for these numbers?
For that matter, why do simple cosmological models based on precise
homogeneity and isotropy fit so well? A completely chaotic and
irregular universe would at first sight seem more probable. If there
is an answer to these questions, it lies in the initial instants of
cosmic history.

For the last 20 years, cosmologists have suspected that the uniformity
is a legacy of something remarkable that happened in the ultraearly
universe: A very fierce cosmical repulsion, it is claimed, could have
accelerated the expansion, so that a tiny patch of space-time expanded
exponentially and homogenized when it was, perhaps, no more than 10-35
seconds old.

The generic idea that our universe inflated from something microscopic
is compellingly attractive. Rather than assuming the expansion as an
initial condition, it accounts for it physically. It looks like
``something for nothing," but it isn't really. That's because our
present vast universe may, in a sense, have zero net energy. Every
atom has an energy because of its mass (Einstein's $mc^2$). But it has a
negative energy due to gravity. We, for instance, are in a state of
lower energy on Earth's surface than if we were up in space. And if we
added up the negative potential energy we possess due to the
gravitational field of everything else, it could cancel out our rest
mass energy. Thus it doesn't, as it were, cost anything to expand the
mass and energy in our universe.

Alan Guth spelled out this concept of inflation in 1981, building on
work by others. It invokes extreme, untested physics and so still has
unsure foundations. But it isn't just metaphysics. Its one generic
prediction -- that the universe would be stretched flat -- seems to be borne
out by observation. Moreover, observations can in principle firm it
up. For instance, the slight ripples that are the seeds of galaxies
and clusters could have been quantum fluctuations, imprinted when the
entire universe was of microscopic size and stretched by inflationary
expansion. The nonuniformities in the present universe depend, in a
calculable way, on the physics of inflation, so observations will be
able to probe this extreme physics and perhaps help us understand what
caused inflation.

Some variants of the inflationary universe-espoused by Andrei Linde,
Alex Vilenkin, and others-suggest that our big bang wasn't the only
one. This fantastic speculation dramatically enlarges our concept of
reality. The entire history of our universe might just be an episode,
one facet, of the infinite multiverse. Current ideas on superstring
theory, and the possibility of extra spatial dimensions, suggest other
equally fascinating scenarios.

\section {   The Universe as We Are Coming to Know It} 

Cosmologists are no longer
starved of data: Current progress is owed far more to observers and
experimentalists than to armchair theorists. Actual physical probes
remain confined to our own solar system, but vicarious exploration
with telescopes and other techniques has extended our horizons. These
tools allow us to study galaxies whose light has been journeying
toward us for 90\% of the time since the big bang.

And we are making sense of what we see. Stars follow finite life
cycles whose main features are now quite well understood. There is no
equivalent understanding of galaxies. But observations of nearby
galaxies, and ones so remote that they are being viewed as they form,
are helping.

We're witnessing a crescendo of discoveries that promises to continue
throughout the next decade. It is something of a coincidence -- of
technology and funding, as well as of the way the intellectual
discourse has developed -- that there has, more or less concurrently,
been an impetus on several fronts: 

\begin{itemize} 
\item Sensitive tools and techniques
make it possible to intensively study the microwave background
fluctuations. 

\item The Hubble Space Telescope (HST) has been
fulfilling its potential for observing deep-space phenomena; new 8- to
10-meter telescopes are on line; new x-ray telescopes in space, and
radio arrays on the ground, now offer greater sensitivity. And a
decade from now, new space telescopes should carry the enterprise
beyond what the HST can achieve.  

\item Large-scale clustering and
dynamics studies, and big surveys of galaxies, will permit sensitive
statistical tests that should discriminate among theories for
structure formation.  

\item Dramatic advances in computer technology
have allowed increasingly elaborate numerical simulations, which now
incorporate realistic gas dynamics as well as gravity.  

\item New
fundamental physics offers the hope of putting the ultraearly universe
on as firm a footing as the later eras.  
\end{itemize}

Some debates have been
settled; some earlier issues are no longer controversial. As the
consensus advances, new questions, which couldn't even have been posed
earlier, are now being debated. Among them are: Why does our universe
consist of the particular mix of baryonic matter and dark matter? What
caused the initial favoritism for matter over antimatter? Is the dark
matter composed of neutral particles surviving from the big bang, or
is it something even more exotic? Are they the outcome of quantum
fluctuations imprinted when our entire universe was of microscopic
size? What is the mysterious dark energy that makes our universe flat,
and how does this relate to inflation?  

It will remain a challenge to
understand the very beginning. This will require a new theory, perhaps
a variant of superstrings, which unifies cosmos and
microworld. Optimists hope for breakthroughs soon. But the aim of
cosmology and astronomy is to map out how a simple fireball evolved,
over 12 billion years, into the complex cosmic habitat we find around
us. Understanding how the consequences of the basic laws have unfolded
over cosmic history is an inexhaustible challenge for the new
millennium.

\section*{Reading}

\def\ref1{\par\noindent\hangafter=1\hangindent=20pt}
\ref1 N. A. Bahcall, J. P. Ostriker, S. Perlmutter,
P. J. Steinhardt, Science 284, 1481 (1999).  
\ref1 M. C. Begelman and
M. J. Rees, Gravity's Fatal Attraction: Black Holes in the Universe
(W. H. Freeman \& Company, New York, 1998).  
\ref1 R. Brawer and
A. P. Lightman, Origins: The Lives and Worlds of Modern Cosmologists
(Harvard University Press, Cambridge, MA, 1992).  
\ref1 E. R. Harrison,
Cosmology: The Science of the Universe (Cambridge University Press,
Cambridge, ed. 2, 2000).  
\ref1 J. A. Peacock, Cosmological Physics (Cambridge University Press, Cambridge, 1998).  
Reviews of Modern
Physics 71(2), (1999).
\ref1 American Physical Society. Special centenary
issue with many authoritative reviews of history and current status of
subject.

\end{document}